\documentstyle[12pt,fleqn]{article}
\input epsf.sty
\begin{document}
\epsfverbosetrue

\newcommand{\be}{\begin{equation}}
\newcommand{\ee}{\end{equation}}
\newcommand{\bea}{\begin{eqnarray}}
\newcommand{\eea}{\end{eqnarray}}
\newcommand{\eps}{\epsilon_{\mu \nu \rho \sigma}}
\newcommand{\ktns}{{\bf k}_T}
\newcommand{\kt}{{\bf k}_T^2}
\newcommand{\ga}{\gamma_{5}}
\newcommand{\gp}{\gamma^{+}}
\newcommand{\smn}{\sigma^{\mu \nu}}
\def\sla#1{\setbox0=\hbox{$#1$}
   \dimen0=\wd0 \setbox1=\hbox{/} \dimen1=\wd1
   \ifdim\dimen0>\dimen1 \rlap{\hbox to \dimen0{\hfil/\hfil}} #1
   \else  \rlap{\hbox to \dimen1{\hfil$#1$\hfil}} / \fi}

\title{\bf
Modelling distribution functions\\ and fragmentation functions
}

\vspace{1 cm}
\author{
J. Rodrigues$^{1,2}$, A. Henneman$^1$ and P.J. Mulders$^{1,3}$\\
\mbox{}\\
$^1$National Institute for Nuclear Physics and High--Energy
Physics (NIKHEF)\\
P.O. Box 41882, NL-1009 DB Amsterdam, the Netherlands\\
and\\
$^2$Department of Physics, Instituto Superior T\'{e}cnico, \\
Av. Rovisco Pais, 1096 Lisboa Codex, Portugal\\
and\\
$^3$Department of Physics and Astronomy, Free University \\
De Boelelaan 1081, NL-1081 HV Amsterdam, the Netherlands
}
\date{}
\maketitle

\vspace{8 cm}
\noindent
October 1995\\
NIKHEF 95-056\\
nucl-th/9510036

\vspace{3 cm}
\noindent
Talk presented at the ELFE Summer School on Confinement Physics,
22-28 July 1995, Cambridge, England

\newpage
\setcounter{page}{1}

\mbox{}
\vspace{3 cm}

\begin{center}

{\Large\bf MODELLING DISTRIBUTION FUNCTIONS \\
\vspace{0.5 cm}
AND FRAGMENTATION FUNCTIONS}

\vspace{1 cm}

J. Rodrigues$^{1,2}$, A. Henneman$^1$ and P.J. Mulders$^{1,3}$
\\
\vspace{.5 cm}
$^1$
{\it National Institute for Nuclear Physics and High--Energy Physics (NIKHEF)\\
P.O. Box 41882, NL-1009 DB Amsterdam, the Netherlands}\\
$^2$
{\it Department of Physics, Instituto Superior T\'{e}cnico, \\
Av. Rovisco Pais, 1096 Lisboa Codex, Portugal}\\
$^3$
{\it Department of Physics and Astronomy, Free University \\
De Boelelaan 1081, NL-1081 HV Amsterdam, the Netherlands} \\

\end{center}

\date{}


\begin{abstract}

We present examples for the calculation of the distribution and
fragmentation functions using
the representation in terms of non-local matrix elements of quark field
operators. As specific examples, we use a simple spectator model to estimate
the leading twist quark distribution functions and the
fragmentation functions for a quark into a nucleon or a pion.

\end{abstract}


\section{Introduction}

{}From the theoretical point of view, the distribution functions are
quark correlation functions involving matrix elements of bilocal combinations
of quark fields \cite{piet1}.
In fact, using the formalism of light-cone quantization, we can give
nice physical interpretations for the leading twist functions
$f_1(x)$, $g_1(x)$ and $h_1(x)$. The first one represents the probability
of finding a quark with (light-cone) momentum fraction $x$. The second
one measures the probability of finding right-handed quarks with momentum
fraction $x$ minus the probability of finding left-handed quarks with
the same momentum fraction. It is called a chirality distribution or
(if we work in the infinite momentum frame) helicity distribution.
In this talk we shall not consider the third one.
A similar interpretation can be given for the fragmentation functions
$D_1(z)$ and $G_1(z)$.

The stucture functions that appear in the cross sections for hard scattering
processes are weighted sums of these (and other) distribution functions.
However, due to the non-perturbative nature of QCD at low energies,
we are not able to calculate
from first principles the distribution functions
nor the fragmentation functions of quarks into hadrons.

We are then lead to build models for the quark correlation functions. In
this work we use a spectator model with some particular vertices that
describe the interaction between the struck quark, the spectactor and
the hadron. Our purpose is to calculate the leading twist distribution
functions and fragmentation functions for the nucleon and the pion.

We are going to neglect the contributions of gluons.
They lead to $\alpha_s$ and $\alpha_s \log Q^2$ corrections in the
structure functions, the latter usually being absorbed in a scale
dependence of the distribution and fragmentation functions. Thus, no QCD
corrections will be computed. In this approximation the structure functions,
the distributions and the fragmentation functions do not depend on $Q^2$
(for fixed $x \equiv -q^+/P^+$ or $z \equiv P_h^-/k^-$).

\section{Distribution Functions}

It is well known that the cross-section
for an inclusive deep inelastic lepton-hadron scattering
can be written in terms of the hadronic tensor

\be \label{ht1}
W_{\mu \nu} = {1 \over 4M} \ \int {d^4\xi \over 2 \pi} e^{iq\xi}
\langle P, S | [J_\mu (\xi), J_\nu (0)] |P, S \rangle ,
\ee

\noindent where $M$ is the mass of the hadron, $P^\mu$ is its four-momentum,
$S^\mu$ is its spin, $q^\mu$ is the momentum carried by the virtual photon
and
 $J_\mu$ is the electomagnetic  current. Since the hadron
wave function is not known, we cannot calculate exactly the matrix element
in ({\ref{ht1}). However, $W_{\mu \nu}$ admits the standard
parametrization in terms of the dimensionless
structure functions $F_1$, $F_2$, $G_1$ and $G_2$ (see, for instance,
ref. \cite{roberts}).

The leading contribution in $W_{\mu \nu}$ can be calculated from the
well known handbag diagram.
This diagram is composed of a
hard scattering part (that can be calculated with perturbative QCD) and
a soft hadronic part which must be
modelled, since it involves unknown physics. This
is the quark correlation function

\bea \label{cf1}
& & \Phi_{ij} (k,P,S) =  \int {d^4 \xi \over (2 \pi)^4} \ e^{i k \cdot \xi} \
\langle P,S \mid \bar{\psi}_j (0) \ \psi_i (\xi) \mid P,S \rangle \nonumber \\
& = & {1 \over (2 \pi)^3} \sum_X \
\langle P,S \mid \bar{\psi}_j (0) \mid X \rangle
 \ \theta (P_{X}^0)
\ \delta \left( (k-P)^2 - M_{X}^2 \right) \
 \langle X \mid \psi_i (0) \mid P,S \rangle ,
\eea

\noindent where $\{ X \}$ is  a complete set of intermediate states.
A similar definition can be given for the antiquark correlation function
$\Phi^c.$

Contained in  this correlation function, is a set of distribution
functions defined by

\be
\Phi^{[\Gamma]} (x) \equiv
\left. {1 \over 2} \ \int dk^- d{\bf k}_T \
Tr (\Phi \Gamma) \right|_{k^+=xP^+}
\ee

\noindent where $\Gamma$ is any $4 \times 4$ matrix.
The leading twist distributions are

\bea
f_1(x) & \equiv & \Phi^{[\gamma^+]} (x) =
{1 \over 2} \ \int dk^- d{\bf k}_T \ Tr (\Phi \ \gamma^+) \\
g_1(x) & \equiv & \Phi^{[\gamma^+ \gamma_5]} (x) =
{1 \over 2} \ \int dk^- d{\bf k}_T \ Tr (\Phi \ \gamma^+ \gamma_5)
\eea

\noindent and the structure functions are given by

\bea
F_2 (x)/x & = & 2 F_1(x) = \sum_f Q_f^2 \ \left[ \Phi^{[\gamma^+]} (x)
+ {\Phi^c}^{[\gamma^+]} (x) \right] \\
2 \lambda \ G_1 (x) & = &
\sum_f Q_f^2 \ \left[ \Phi^{[\gamma^+ \ga]} (x)
+ {\Phi^c}^{[\gamma^+ \ga ]} (x) \right] ,
\eea

\noindent $\lambda$ being the helicity of the hadron.
Moreover, it can be shown that $G_2$ enters in the
cross section multiplied by a factor $1/Q$ (it is a twist 3 structure
function) and then we shall neglect it.

The most general expression for $\Phi$ is  \cite{ralston1}

\bea \label{cf3}
\Phi(k,P,S) & = & M \ A_{1} +  A_{2} \sla P + A_{3} \ \sla k +
{A_{4} \over M} \ \sigma^{\mu \nu} P_{\mu} k_{\nu} + i \ A_5 \ k
\cdot S \ \ga +  M \ A_{6} \ \sla S \ \ga \nonumber \\ & & + \
{A_{7} \over M} \ k \cdot S \ \sla P \ \ga +
{A_{8} \over M} \ k \cdot S \ \sla k \ \ga +
\ i \ A_{9} \ \smn \ \ga \ S_{\mu} P_{\nu} \\ & & + \
i \ A_{10} \ \smn \ \ga \ S_{\mu} k_{\nu} + \
i \ {A_{11} \over M^2} \ k \cdot S \ \smn \ \ga \ k_{\mu} P_{\nu}
+ {A_{12} \over M} \ \eps \ \gamma^{\mu} P^{\nu} k^{\rho} S^{\sigma}.
\nonumber
\eea

\noindent
Hermiticity requires that all the amplitudes $A_i (k \cdot P, k^2)$ are real.
When time-reversal invariance applies, the amplitudes $A_4$, $A_6$
and $A_{12}$ vanish.

The distributions can be written in terms of these amplitudes \cite{rik94p7}:

\bea \label{dist1}
f_1(x) & = & M^4 \ \pi \int d\sigma d\tau \ \theta (x\sigma - \tau - x^2) \
(A_2 + xA_3) \\ \label{dist2}
g_1(x) & = & M^4 \ \pi \int d\sigma d\tau \
\theta (x\sigma - \tau - x^2)
 \left[ -A_6 - \left( {\sigma \over 2} -x \right) \ (A_7+xA_8) \right]
\eea

\noindent with $\tau = k^2/M^2, \sigma = 2 k \cdot P /M^2.$

\section{Fragmentation Functions}

Also the calculation of the fragmentation functions
involves a soft hadronic part given by

\be \label{ff1}
\Delta_{ij} (k, P_h, S_h) = \sum_X \ \int {d^4 \xi \over {(2 \pi)}^4} \
e^{ik \xi} \ \langle 0 | \psi_i (\xi) | P_h, S_h, X \rangle \
\langle X, P_h, S_h | \bar{\psi}_j(0) | 0 \rangle.
\ee

\noindent This function admits an expansion similar to (\ref{cf3}) but we
are not going to need it. Instead, we define immediately the set of functions

\be \label{ff2}
\Delta^{[\Gamma]}(z) \equiv \left. {1 \over 4z} \
\int dk^+ d^2 {\bf P}_{hT} \
Tr (\Delta \Gamma) \right|_{k^-=P_h^-/z}
\ee

\noindent where $z$ is the (light-cone) momentum fraction carried by
the observed hadron originated by the fragmentation of the quark and
${\bf P}_{hT}$ is the transverse momentum of this hadron in a frame where
the hadron parent has no transverse momentum.

The leading twist fragmentation functions are

\bea \label{ff3a,ff3b}
D_1(z) & \equiv & \Delta^{[\gamma^-]}(z) \\
G_1(z) & \equiv & \Delta^{[\gamma^- \ga]}(z)
\eea

\noindent and we can easily derive a useful
relation between these fragmentation functions
and  the previously defined distribution functions:

\bea \label{ff4a}
D_1(z) & = & z/2 \ f_1 (1/z) \\ \label{ff4b}
G_1(z) & = & z/2 \ g_1 (1/z) .
\eea

\section{The Spectator Model}

In the spectator model the sum over all possible intermediate states
in (\ref{cf1}) is reduced to a single term, corresponding to the
spectator $|X_{sp} \rangle $ . It seems reasonable to consider
this spectator as a (scalar or axial vector) diquark.
The amplitude for obtaining the state $|X_{sp} \rangle,$ after removing
a quark from the nucleon is given by

\be \label{sp1}
\langle X_{sp} \mid \psi_i(0) \mid P, S \rangle
= {\left[ {i \over \sla k - m + i \epsilon}  \
\Gamma^{s,a} \ u (P,S) \right]}_i .
\ee

We are going to assume that the vertices have the following structure:

\bea \label{sp2a,sp2b}
\Gamma^s & = & g_s(k^2) \ I \\
\Gamma_\alpha^a & = &
{g_a(k^2) \over \sqrt 3} \ \left( \gamma_\alpha \ \gamma_5
-{P_\alpha \over M} \ \gamma_5 \right) .
\eea

As form factors, we follow ref. \cite{thomas} and choose

\be \label{sp3}
g_{s,a}(k^2) = N_{s,a} {\left( M^2 \right)}^{\alpha-1} \
{k^2 - m^2 \over {(k^2 - \Lambda^2)}^\alpha} .
\ee

Throughout this section, $m$ stands for the (effective)
quark mass, $\Lambda$ is a parameter to adjust and
$N$ is a normalization constant. The other parameter of the model is the
mass of the diquark. In order to fit the experimental data as well as
possible, we set $m=0.36$ GeV, $m_s=0.6$ GeV, $m_a = 0.8$ GeV
and $\Lambda=0.7$ GeV.

The choice $\alpha = 2$ in (\ref{sp3}) ensures the correct behaviour of
the distributions at large $x.$ For the case of
a scalar spectator the correlation function is then

\bea \label{sp4}
\Phi^s & = & {\theta (P_{s}^0) \over 16 \pi^3} \ {\delta ((k-P)^2 - m_{s}^2)
\over (k^2-m^2)^2} \times \nonumber \\ & & [(\sla k +m) \ \Gamma^s(k,P) \
(\sla P + M) \ (1 + \ga \sla S) \ \gamma_0 \
{\Gamma^s}^\dagger (k,P) \ \gamma_0 \ (\sla k + m)]
\eea

\noindent where $m_s$ is the mass of the scalar diquark.
A similar expression is valid for the case of a vector diquark.

Using equations (\ref{sp1}) - (\ref{sp4}) we can readily compute the
amplitudes $A_i$ of (\ref{cf3}) and then calculate the distributions
using (\ref{dist1}) and (\ref{dist2})
and the fragmentation functions using (\ref{ff4a}) and (\ref{ff4b}).
A calculation of fragmentation functions for baryons has been given in
Ref. \cite{hoodbhoy}.

For the pion one can also employ the spectator model. The spectator is now an
antiquark. The correlation function can be written in terms of only
four amplitudes (because the pion is a spinless particle).
The vertex is now

\be
\Gamma^\pi = {g(k^2) \over {\sqrt 2}} \ \ga \ \left( \sla P + M \right)
\ee

\noindent with the same form factor as in (\ref{sp3}) but with $\alpha = 3/2$
and $\Lambda = 0.4$ GeV.

The preliminary results obtained with this model are shown in Fig. 3.
At this moment no comparison with experiment is possible. This requires
QCD evolution of these functions, what we postpone for a later work.
At this stage the results ilustrate how the framework of quark correlation
functions enables model calculations for distribution {\em and} fragmentation
functions.
\begin{figure}[hhh]
\vspace{-2.5cm}
\label{graficos}
\begin{center}
\leavevmode
\epsfxsize=14 cm
\epsfbox{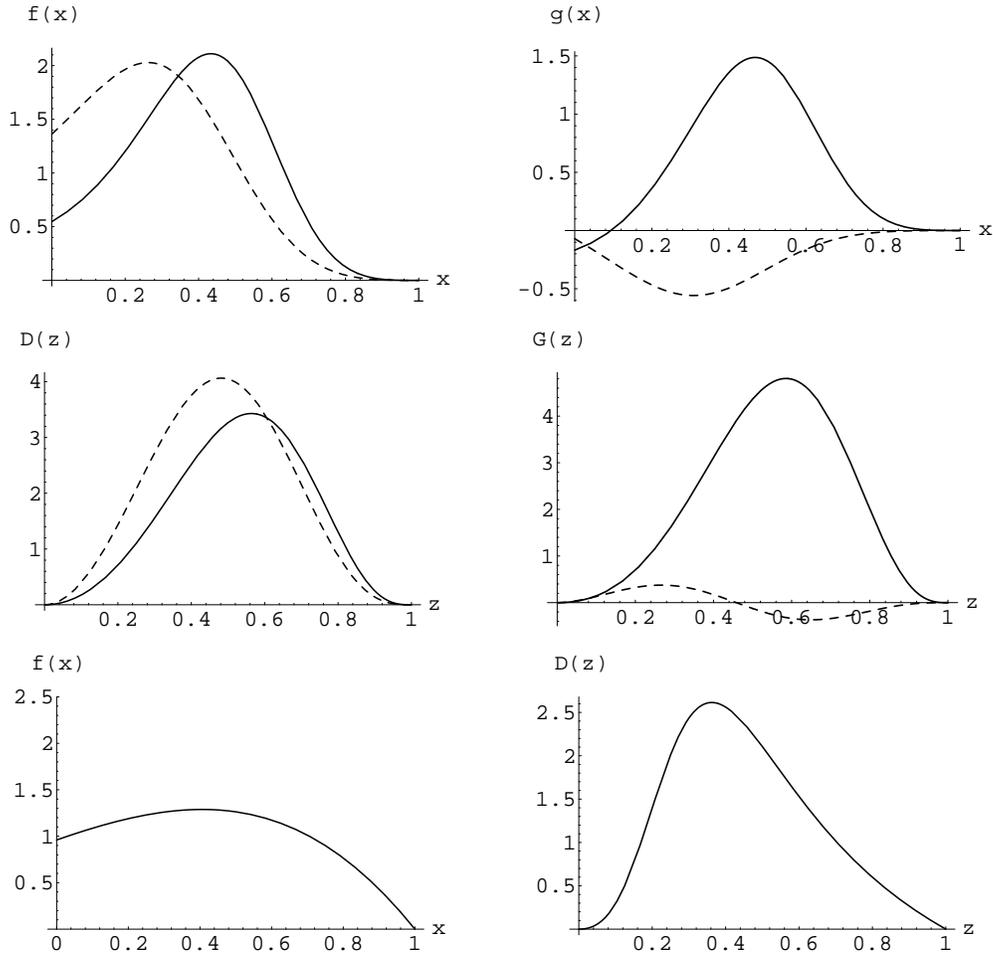}
\end{center}
\vspace{-3cm}
\caption{Nucleon distributions $f_1(x)$ and $g_1(x)$ (top);
nucleon fragmentation functions $D_1(z)$ and $G_1(z)$ (middle);
pion distribution function $f_1(x)$ (bottom-left);
pion fragmentation function $D_1(z)$ (bottom-right).
For the nucleon, the full line is for the scalar diquark and the
dashed line for the vector diquark.
The normalization of the fragmentation functions is arbitrary.}
\end{figure}

This work is part of the research program of the foundation for Fundamental
Research of Matter (FOM) and the National Organization for Scientific Research.

\end{document}